\pdfoutput=1
\documentclass[twocolumn,superscriptaddress,aps,preprintnumbers,amsmath,amssymb,prl,nofootinbib]{revtex4-1}
\usepackage{graphicx}
\usepackage{bm}
\usepackage{color}
\usepackage{amsmath,amssymb}	
\usepackage[colorlinks=true, linkcolor=blue, citecolor=blue,urlcolor=black]{hyperref} 
\usepackage{setspace}
\usepackage{longtable}
\usepackage{booktabs}
\usepackage{comment}
\usepackage{array}
\usepackage{cancel}
\usepackage{enumitem}
\usepackage{todonotes}
\usepackage{physics}
\def\slashchar#1{\setbox0=\hbox{$#1$} 
\dimen0=\wd0 
\setbox1=\hbox{/} \dimen1=\wd1 
\ifdim\dimen0>\dimen1 
\rlap{\hbox to \dimen0{\hfil/\hfil}} 
#1 
\else 
\rlap{\hbox to \dimen1{\hfil$#1$\hfil}} 
/ 
\fi}

\newcommand{\csunit}{\mathrm{cm}^2/\mathrm{g}}

\begin{document}
\newcolumntype{Y}{>{\centering\arraybackslash}p{23pt}} 


\preprint{IPMU25-0013}

\title{Stringent Constraints on Self-Interacting Dark Matter\\ Using Milky-Way Satellite Galaxies Kinematics}

\author{Shin'ichiro Ando}
\affiliation{GRAPPA Institute, University of Amsterdam, Science Park 904, 1098 XH Amsterdam, The Netherlands}
\affiliation{Kavli Institute for the Physics and Mathematics of the Universe (WPI), \\The University of Tokyo Institutes for Advanced Study, \\ The University of Tokyo, Kashiwa 277-8583, Japan}

\author{Kohei Hayashi}
\affiliation{National Institute of Technology, Sendai College, Natori, Miyagi 981-1239, Japan}
\affiliation{Astronomical Institute, Tohoku University, Aoba-ku, Sendai 980-8578, Japan}
\affiliation{ICRR, The University of Tokyo, Kashiwa, Chiba 277-8582, Japan}

\author{Shunichi Horigome}
\affiliation{Kavli Institute for the Physics and Mathematics of the Universe (WPI), \\The University of Tokyo Institutes for Advanced Study, \\ The University of Tokyo, Kashiwa 277-8583, Japan}

\author{Masahiro Ibe}
\affiliation{ICRR, The University of Tokyo, Kashiwa, Chiba 277-8582, Japan}
\affiliation{Kavli Institute for the Physics and Mathematics of the Universe (WPI), \\The University of Tokyo Institutes for Advanced Study, \\ The University of Tokyo, Kashiwa 277-8583, Japan}

\author{Satoshi Shirai}
\affiliation{Kavli Institute for the Physics and Mathematics of the Universe 
(WPI), \\The University of Tokyo Institutes for Advanced Study, \\ The University of Tokyo, Kashiwa 277-8583, Japan}

\date{\today}
\begin{abstract}
Self-interacting dark matter (SIDM) has been proposed to address small-scale challenges faced by the cold dark matter (CDM) paradigm, such as the diverse density profiles observed in dwarf galaxies. In this study, we analyze the kinematics of dwarf galaxies by incorporating the effects of gravothermal core collapse into SIDM models using a semi-analytical subhalo framework. Our analysis covers the stellar kinematics of both classical and ultrafaint dwarf galaxies. 
The results indicate a bimodal preference for small and large self-interaction cross sections in ultrafaint dwarf galaxies, while in classical dwarfs, larger cross sections progressively decrease the model's statistical support. The combined analysis decisively prefers CDM to SIDM when the self-interaction cross section per unit mass, $\sigma/m$, exceeds $\sim$0.2$\,\csunit$,
if a velocity-independent cross section is assumed.
Our study significantly enhances our understanding of dark matter dynamics on small scales.
\end{abstract}

\maketitle

\textbf{\textit{Introduction}}---The collisionless cold dark matter (CDM) paradigm successfully explains the large-scale structure of the Universe (e.g.,~\cite{Vogelsberger:2014kha,Vogelsberger:2014dza,Schaye:2014tpa,Springel:2017tpz}), but faces challenges on small scales (for reviews, see Refs.\,\cite{Tulin:2017ara,Bullock:2017xww}). 
For instance, the cored dark matter density profiles observed in dwarf spheroidal galaxies (dSphs) contradict the the cuspy profiles predicted by CDM-only $N$-body simulations~\cite{McGaugh_1998,de_Blok_2001,Oh_2011,Read:2017lvq}.
However, recent dynamical studies suggest that whether dSphs exhibit cores or cusps remains debated due to large uncertainties~\cite{Hayashi:2020jze,Hayashi:2022wnw,2022NatAs...6..659B}.

Self-interacting dark matter (SIDM)~\cite{Spergel_2000} has been proposed to address these discrepancies by facilitating heat transport within dark matter halos.
This mechanism leads to the formation of central density cores and, in later stages, gravothermal core collapse~\cite{Balberg:2001qg, Koda:2011yb}, introducing unique dynamics distinct from CDM.
Observational tests of SIDM are therefore of significant interest.

Reference~\cite{Hayashi:2020syu} tested the SIDM using the stellar kinematics of 23 ultrafaint dSphs (UFDs), employing halos with an isothermal core and an outer Navarro-Frenk-White (NFW) profile~\cite{Kaplinghat_2016,Valli:2017ktb}.
The analysis found no evidence favoring a nonzero self-interaction cross section per unit mass, $\sigma/m$, imposing constraints thereon.

\begin{figure}[t!]
    \centering
    \includegraphics[width=\linewidth]{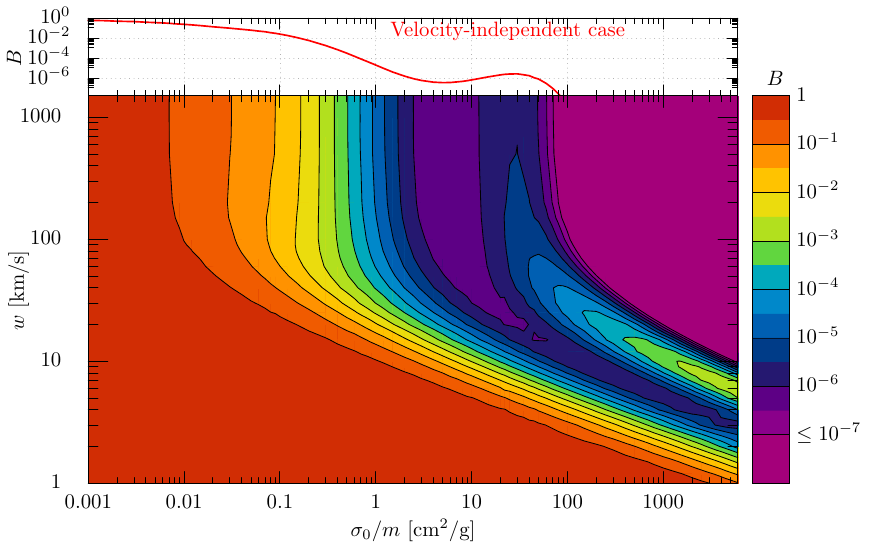}
    \caption{Contour plot of the Bayes factor, $B$, shown as a function of the self-interaction cross section per unit mass, $\sigma_0/m$, and the velocity parameter, $w$ [both defined in Eq.~(\ref{eq:diff_cross_section})]. 
    A positive (negative) $\log_{10} B(\sigma_0/m, w)$ indicates a preference for SIDM (CDM), with $\log_{10} B \leq -2$ typically regarded as decisive evidence against SIDM. The constraints are derived from Bayesian model comparison using the combined kinematic data of classical and ultrafaint dSphs.}
    \label{fig:2D}
\end{figure}

In this paper, we improve SIDM tests by incorporating gravothermal core collapse.
Unlike previous studies (e.g., \cite{Hayashi:2020syu}) that overlooked the impact of core collapse, we include recent insights suggesting tidal interactions accelerate this process~\cite{Nishikawa:2019lsc,Kummer:2019yrb,Robles:2019mfq,Sameie:2019zfo,Kahlhoefer:2019oyt,Nadler:2020ulu, Correa:2020qam}.
To model core collapse, we adopt a semi-analytical subhalo model SASHIMI~\cite{Ando:2024kpk},\footnote{\url{https://github.com/shinichiroando/sashimi-si}} extended to SIDM by incorporating a parametric SIDM halo model~\cite{Yang:2023jwn}, linking SIDM subhalo evolution to their CDM counterparts.
This framework efficiently predicts the SIDM density profiles, including core formation and gravothermal collapse.
SASHIMI enables a comprehensive survey over the SIDM parameter space, previously hindered by computational costs.

We compare these predictions with the kinematic data of 8~classical and 23~UFDs within the Milky Way.  
As illustrated in Fig.\,\ref{fig:2D}, our analysis decisively favors CDM over SIDM for self-interaction cross sections above $\sigma_0/m \sim 0.2\,\csunit$ in the velocity-independent case. This result imposes stringent constraints on SIDM models while underscoring the pivotal influence of gravothermal collapse in shaping the dynamics of dark matter halos.

\textbf{\textit{Semi-analytical SIDM subhalo model}}---For the density profile of 
the SIDM subhalos as a function the radius from its center $r$, we adopt~\cite{Robertson_2016},  
\begin{align}
\label{eq:rhoc}
    \rho(r) = \frac{\rho_sr_s^3  }{  (r^4 + r_c^4)^{1/4} (r+r_s)^2 } \ ,
\end{align}
up to the the tidal truncation radius $r_t$ beyond which
the density decreases quickly~\cite{Yang:2023jwn}.
Here, $\rho_s$ and $r_s$ are scale density
and radius, respectively.
Within the core size $r_c$, 
the density profile is flat.
The CDM subhalo profile corresponds to that with $r_c=0$.

Throughout this paper, we consider the following differential cross section characterized by two parameters $(\sigma_0, w)$\,\cite{Yang:2022hkm},
\begin{align}
\frac{d\sigma}{d\cos\theta} = \frac{\sigma_0}{
2 \qty[1+(v/w)^2 \sin^2(\theta/2)]^2
}\ ,
\label{eq:diff_cross_section}
\end{align}
where, $\sigma_0$ denotes the nominal cross section, $\theta$ represents the scattering angle, $v$ is the relative velocity between dark matter particles, and $w$ is the ratio of the mediator mass of the self-interaction to the dark matter mass.

The SIDM subhalos 
follow self-similar time evolution~\cite{Balberg:2001qg,Koda:2011yb,Pollack:2014rja,Essig:2018pzq,Outmezguine:2022,Zhong:2023} which is universally given by functions of $\tilde{t}=(t-t_f)/t_c$ with $t_f$ being the halo formation time.
The collapse timescale $t_c$ is obtained via~\cite{Pollack:2014rja,Balberg:2001qg}
\begin{align}
t_c =  \frac{150}{C\qty(\sigma_\mathrm{eff}/m) \hat{\rho}_s \hat{r}_s} \frac{1}{
\sqrt{4\pi G \hat{\rho}_s}}\ ,
\label{eq:core_collapse_timescale}
\end{align}
where $C = 0.75$~\cite{Koda:2011yb,Essig:2018pzq,Nishikawa:2019lsc,Yang:2022zkd} and $G$ is the Newton's gravitational constant. 
Hereafter, parameters with a hat represent those of the CDM model.
For the subhalo with $\tilde{t}>1$ at present, the core collapse phase has been reached.
In our analysis, we terminate the SIDM effect at $\tilde{t}=1.1$, 
which only affects the deeply collapsed central density.

The prior distributions of the SIDM subhalo parameters $(\rho_s,r_{s},r_{c},r_{t})$
are obtained by SASHIMI-SIDM
for a given set of cross section parameters $(\sigma_0/m,w)$.
We first generate a list of time evolutions of the CDM
subhalo parameters of the truncated NFW profile  $\qty(\hat{\rho}_s,\hat{r}_s,r_t)$ 
between the redshifts at the onset of the accretion $z_a$ and at present, i.e., $z=0$ by SASHIMI\,\cite{Hiroshima:2018kfv,Ando:2019xlm}.
For each realization of  $(\hat{\rho}_s, \hat{r}_s)$, 
we obtain $(\hat{V}_\mathrm{max}, \hat{r}_\mathrm{max})$
via $\hat{V}_{\mathrm{max}}=1.64\,\hat{r}_s\sqrt{G
\hat{\rho}_s}$ and $\hat{r}_{\rm max} = 2.163\,r_s$.

Once we have the evolution of $(\hat{V}_\mathrm{max}, \hat{r}_\mathrm{max},t_c)$
for a given realization,
we obtain $(V_\mathrm{max},r_\mathrm{max})$
of the SIDM at present by the 
\textit{integral approach} 
in Ref.\,\cite{Yang:2023jwn}.
Finally, from $(V_\mathrm{max},r_\mathrm{max})$
of the SIDM at present, we obtain the SIDM subhalo
parameters $(\rho_s,r_s,r_c,r_t)$ through the 
\textit{basic approach} 
in Ref.\,\cite{Yang:2023jwn}. 
The conversion between the CDM subhalo
parameters and the SIDM subhalo parameters
in Ref.\,\cite{Yang:2023jwn}
are calibrated by $N$-body SIDM simulation in Ref.\,\cite{Yang:2022hkm} and validated by the cosmological zoom-in SIDM simulation\,\cite{Yang:2022mxl}.

\begin{figure}[t!]
    \centering
    \includegraphics[width=\linewidth]{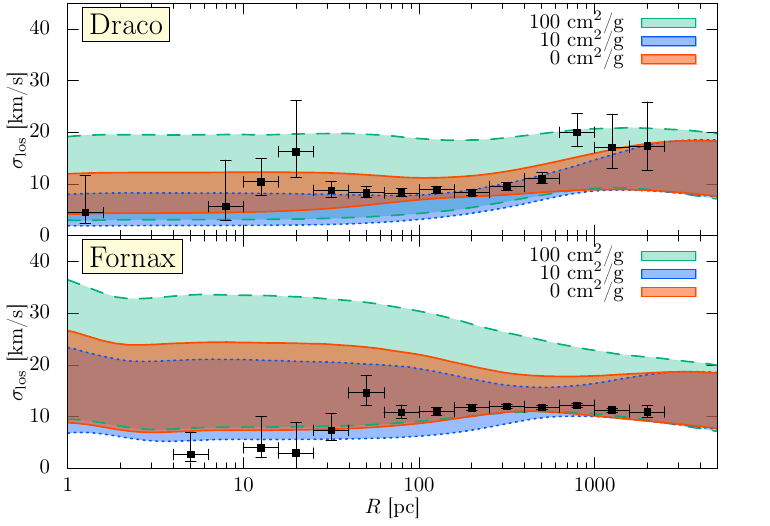}
    \caption{The 68\% credible intervals of the SASHIMI-SIDM prior distributions for the line-of-sight velocity dispersion profiles, $\sigma_{\rm los}(R)$, of the classical dSph Draco (top) and Fornax (bottom). The predictions are shown as colored bands corresponding to different values of the velocity-independent self-interaction cross-section, $\sigma/m$. Observed velocity dispersion data are overlaid as points with 1-$\sigma$ error bars.}
    \label{fig:sigmalos_classical}
\end{figure}

\begin{figure}[t]
    \centering    
    \includegraphics[width=0.95\linewidth]{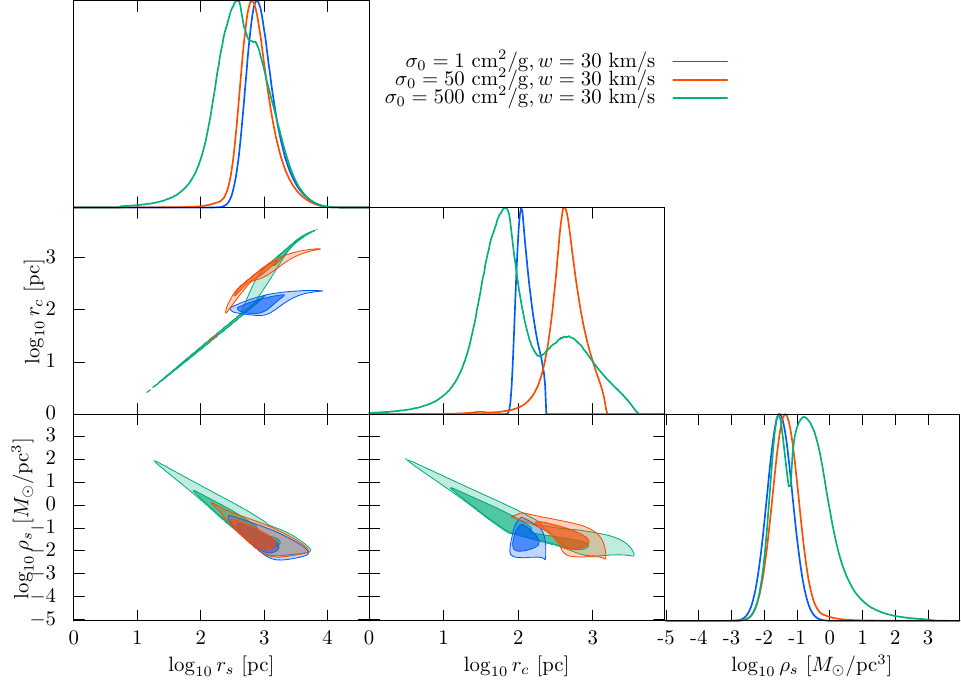}
    \caption{Corner plot showing the SIDM parameter distributions, $P_{\rm sat}(\rho_s,r_s,r_c)$, predicted by SASHIMI-SIDM for subhalos that could host UFDs ($V_{50} = 10.5\,\mathrm{km/s}$ ). Different colors corresponds to the different cross-section parameter values: $\sigma_0 = 1\,\csunit$ (blue), $\sigma_0 = 50\,\csunit$ (red), and $\sigma_0 = 500\,\csunit$ (green), with $w = 30\,\mathrm{km/s}$. The off-diagonal 2D plots indicate the 68\% and 95\% Bayesian credible intervals for each nominal cross-section.
    }
    \label{fig:prior}
\end{figure}

\textbf{\textit{Satellite prior}}---Based on the prior distribution of SIDM subhalo parameters, $P_{\rm sh}(\rho_s,r_s,r_c|\sigma_0,w)$, we shall construct the \textit{satellite prior}\,\cite{Ando:2020yyk, Horigome:2022gge} (see Fig.\,\ref{fig:prior}).
We assume that a satellite galaxy forms in its host subhalo with a probability $P_{\rm form}$ that depends on the maximum circular velocity at its peak, $V_{\rm peak}$; in the semi-analytical model, it is realized at accretion, i.e., $V_{\rm peak} = V_{\rm max}(z_a)$.
The satellite prior is therefore
\begin{align}
    &P_{\rm sat}(\rho_s,r_s,r_c|\sigma_0,w)\cr
    &\propto \!\!\ P_{\rm sh}(\rho_s,r_s,r_c|\sigma_0,w)P_{\rm form}(V_{\rm peak}) \ ,
\end{align}
for the latter term of which we adopt
\begin{align}
P_{\rm form}(V_{\rm peak}) = \frac{1}{2}\left[1+\mathrm{erf}\left(\frac{V_{\rm peak}-V_{\rm 50}}{\sqrt{2}\sigma_\mathrm{peak}}\right)\right] \ ,
\end{align}
where $V_{50}$ is a parameter at which $P_{\rm form} = 0.5$.
As a canonical parameter choice, we adopt $V_{50} = 25 \, \mathrm{km/s}$ and $\sigma_\mathrm{peak} = 0$ for classical dSphs (step function), and $V_{50} = 10.5 \,\mathrm{km/s}$ and $\sigma_\mathrm{peak} = 2.5\, \mathrm{km/s}$ for UFDs, respectively.
See Refs.\,\cite{Ando:2020yyk, Horigome:2022gge} for a more detail.

\begin{figure}[t]
    \centering
    \includegraphics[width=0.95\linewidth]{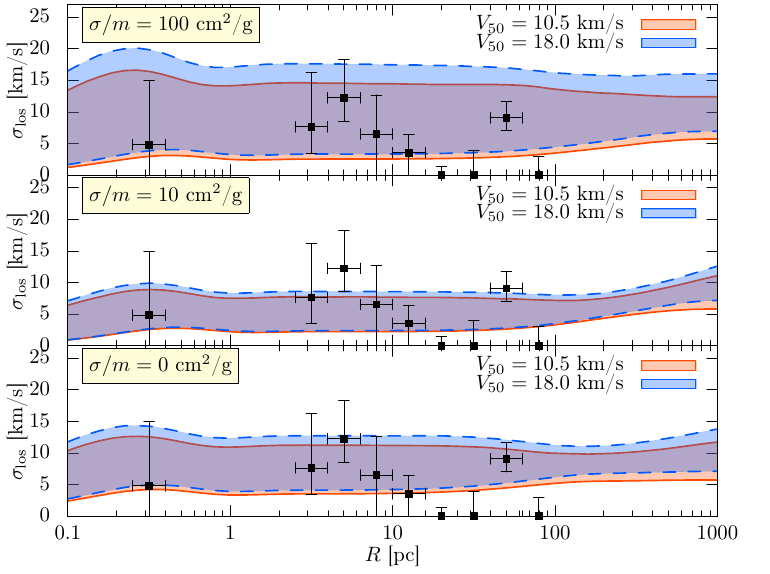}
    \caption{The comparison of the $\sigma_\mathrm{los}$ profiles from the SASHIMI-SIDM prior distributions and the observational data for Segue~1. The red and blue bands represent the priors based on $V_{50}=10.5$~km/s and $18.0$~km/s, respectively.}
    \label{fig:sigmalos_seg1}
\end{figure}

In Fig.\,\ref{fig:sigmalos_classical}, we present the prior distributions of the line-of-sight velocity dispersion profile, $\sigma_{\rm los}(R)$ [defined later in Eq.\,(\ref{eq:losdisp})], for the Draco and Fornax classical dSphs, considering various values of the self-interaction cross section $\sigma_0/m$. (Profiles for UFDs, Segue~1, are shown in Fig.\,\ref{fig:sigmalos_seg1}.)
For $\sigma_0/m = 10\,\csunit$, the reduction in $\sigma_{\rm los}$ is attributed to core formation, whereas for $\sigma_0/m = 100\,\csunit$, the increase in $\sigma_{\rm los}$ results from gravothermal collapse. 
Notably, the observed $\sigma_{\rm los}$ data for Fornax are inconsistent with scenarios involving large self-interaction cross sections, as the predicted velocity dispersion profile does not support the presence of gravothermal collapse~(see also Fig.\,\ref{fig:prior}).

In the following, we utilize $P_{\rm sat}(\rho_s,r_s,r_c|\sigma_0/m,w)$ for each value of $(\sigma_0/m,w)$ as the prior distribution for the Bayesian inference of the SIDM cross section parameters. 
The truncation radius $r_t$ is not constrained by the kinematical data and is therefore irrelevant  as long as $r_c < r_s < r_t$ holds, which we confirm is the case.

\textbf{\textit{Analysis Method}}---The Jeans equation describes the motion of stars in a dark matter-dominated system by relating the stellar velocity dispersion, density, and gravitational potential $\Phi$ under the assumption of dynamical equilibrium~\cite{2008gady.book.....B}. 
Assuming spherical symmetry, the Jeans equation is given by
\begin{align}
\label{eq:Jeans}
   \frac{\partial\nu(r) \sigma_r^2(r)}{\partial r} 
        + 
    \frac{2\beta_{\mathrm{ani}}(r)\sigma_r^2(r)}{r} = - \nu(r) \frac{\partial \Phi}{\partial r},
\end{align}
where $r$ denotes the radius from the center of the dSph and 
$\nu(r)$ is the intrinsic stellar density. 
$\beta_{\mathrm{ani}}(r)\equiv 1-\sigma^2_{\theta}/\sigma^2_r$ is the velocity anisotropy parameter.

Solving Eq.\,\eqref{eq:Jeans}, we obtain the stellar velocity dispersion profile of the radial component, $\sigma_r(r)$.
We convert $\sigma_r(r)$ to the 
line-of-sight averaged dispersion profile
by integrating $\sigma_r(r)$ along the line-of-sight,
\begin{align}
\label{eq:losdisp}
\sigma_{\rm los}^2(R)=\frac{2}{\Sigma(R)}\int_R^\infty\!\! dr \Bigl(1-\beta_{\mathrm{ani}}(r)\frac{R^2}{r^2}\Bigr)\frac{\nu(r)\sigma_r^2(r)}{\sqrt{1-R^2/r^2}}\ .
\end{align}
Here, $R$ is a projected radius from the center of the dSph,  
$\Sigma(R)$ is the stellar surface density profile derived from the intrinsic stellar density $\nu(r)$ through the Abel transform.
Following the previous work~\cite{Hayashi:2020syu}, the stellar density and the anisotropy profiles are given by the Plummer~\cite{Plummer:1911zza} and Baes~\& van Hese~\cite{Baes:2007tx} parametrisations, respectively~(the detailed formulae and their parameters are described in the Appendix).
In this work, we assume that the half-light radius of the Plummer model, $ r_{1/2} $, follows a Gaussian prior distribution with a mean of $ r^{0}_{1/2} $ and a standard deviation of $\delta r_{1/2} $. Here, $ r^{0}_{1/2}$  and $ \delta r_{1/2} $ correspond to the photometrically measured value and its associated uncertainty, respectively.

The likelihood function is constructed by comparing the stellar kinematics data with the predicted velocity dispersion profile, given by:
\begin{align}
    -2 \log({\cal L}) = \sum_{i} \left[ \frac{(v_i -V)^2}{\sigma_{i}^2} + \log(2\pi \sigma_i^2) \right]. \label{eq:like_dis}
\end{align}
Here, $i$ runs the member stars of each dSph with the observed line-of-sight velocity $v_i$, and $V$ is the mean line-of-sight velocity of the member stars.
The dispersion of the line-of-sight velocity $\sigma_{i}^2$ is the squared sum of the line-of-sight velocity dispersion in Eq.\,\eqref{eq:losdisp} and the measurement error $\varepsilon_i$, 
$\sigma_i^2 = \sigma_{\rm los}^2(R_i) + \varepsilon_i^2$.
We always take $V$ to maximize the likelihood ${\cal L}$, i.e., $d \log({\cal L})/dV = 0$.

In the present SIDM model, we have three parameters, $r_s$, $\rho_s$, and $r_c$, representing the dark matter halo density profile, and five astrophysical parameters~(see the Appendix).
For a given set of SIDM halo parameters, the marginal likelihood for each galaxy is given by,
\begin{align}
    \mathcal{L}_\mathrm{eff}(r_s, \rho_s, r_c) = \int d[r_\beta, \eta, \beta_{0,\infty}, r_{1/2}] \mathcal{L} \ ,
\end{align}
where $\int d[r_\beta, \eta, \beta_{0,\infty}, r_{1/2}]$ represents integration over the astrophysical parameters (partly introduced in the Appendix) under the prior distributions.

Once the likelihood function is obtained, the model evidence for a given  $(\sigma_0/m, w)$ can be estimated by combining it with the prior distribution derived from SASHIMI-SIDM. For each dSph, the model evidence is calculated as:
\begin{align}
    &Z(\sigma_0/m, w) \cr
    &= \!\!\int\! d\rho_s d r_{s} d r_c \mathcal{L}_\mathrm{eff}(r_s, \rho_s, r_c) P(\rho_s, r_s, r_c | \sigma_0/m, w) \ . \cr
\end{align}

Finally, we define the Bayes factor of the cross section parameters as $B(\sigma_0/m, w) = Z(\sigma_0/m, w)/Z(0, 0)$,
allowing us to assess the goodness-of-fit of the SIDM model compared to the CDM model.
For the analysis with the combined kinematical data of multiple dSphs, 
the Bayes factor is given by
\begin{align}
    B(\sigma_0/m,w)=\frac{\prod_{k}Z_k(\sigma_0/m,w)}{\prod_{k}Z_k(0,0)}\ ,
\end{align}
where $k$ runs over the dSphs.

For reference, we summarize the Jeffreys' scale for evidence categorization based on the Bayes factor $B$~\cite{Kass:1995loi}. 
For $0 > \log_{10}B\ge-0.5$, 
evidence is barely worth mentioning, 
$-0.5 > \log_{10}B \ge -1$ indicates substantial evidence,
$-1 > \log_{10}B \ge -1.5$, strong evidence,
$-1.5 > \log_{10}B \ge -2$, very strong evidence,
and $\log_{10}B < -2$, decisive evidence favoring the denominator hypothesis. 
For $B \ge 1$, the Bayes factor supports the numerator hypothesis with reversed inequalities (multiplied by $-1$).

\textbf{\textit{Data}}---In this work, we investigate SIDM properties for 8 classical dSphs and 23 UFDs  associated with the Milky Way.
The basic structural properties (the positions of the centers, distances, and half-light radii with the Plummer profile, $r^0_{1/2}$) of their galaxies are adopted from the original observation papers~\cite{2018ApJ...860...66M,2015ApJ...813...44L,2015ApJ...813..109D,2015ApJ...805..130K}.
For the stellar-kinematics of their member stars, we utilize the currently available data taken from each spectroscopic observation paper~\cite{2015MNRAS.448.2717W,2018AJ....156..257S,2016ApJ...830..126F,2009AJ....137.3100W,2008ApJ...675..201M,2007AJ....134..566K,2021ApJ...920...92J,2018MNRAS.480.2609L,2021A&A...651A..80Z,2022ApJ...939...41C,2023AJ....165...55C,2007ApJ...670..313S,2011ApJ...733...46S,2017ApJ...838...11S,2020ApJ...892..137S,2013ApJ...770...16K,2017ApJ...838...83K,2011AJ....142..128W,2015ApJ...811...62K}.
The membership selections for each galaxy follow the methods described in the cited papers.
The unresolved binary stars in a stellar kinematic sample may affect the measured velocity dispersion of our target galaxies due to binary orbital motion.
However, several papers show that binary star candidates can be excluded from the member stars and suggest that such an effect is much smaller than the measurement uncertainty of the velocity.
Therefore, we suppose that the effect of binaries are negligible.

\begin{figure}[t!]
    \centering
    \includegraphics[width=\linewidth]{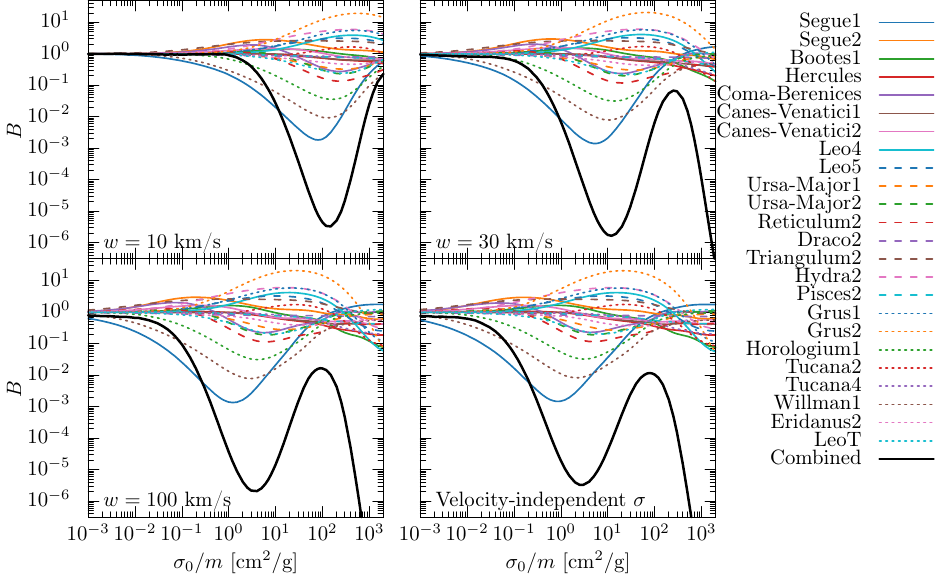}
   \caption{Bayes factor, $B$, for UFDs as a function of $\sigma_0/m$, shown for 
    various values of $w$. Colored (dashed) lines indicate individual galaxies, and the black line shows the combined odds.}
    \label{fig:UFD}
\end{figure}

\textbf{\textit{Results}}---In Fig.\,\ref{fig:UFD}, 
we show the Bayes factor for various UFDs as a function of $\sigma_0/m$, across different values of $w$.
The black solid line represents the combined UFDs odds.
For $w = 10 \, \mathrm{km/s}$ and $w = 30 \, \mathrm{km/s}$, the Bayes factors display bimodal features, with a preference for small ($\sigma_0/m\lesssim 1\,\csunit$) and large cross sections ($\sigma_0/m\gtrsim 10^{2}\text{--}10^{3}\,\csunit$),
with a significant dip at intermediate cross sections ($\sigma_0/m \sim 10\,\csunit$). 
This bimodality reflects the complex interplay between core formation and gravothermal collapse in SIDM models.
For larger $w$, 
which approaches the behavior of a model with a velocity-independent cross section, the Bayes factors consistently show a preference for CDM.

\begin{figure}[t!]
    \centering
    \includegraphics[width=\linewidth]{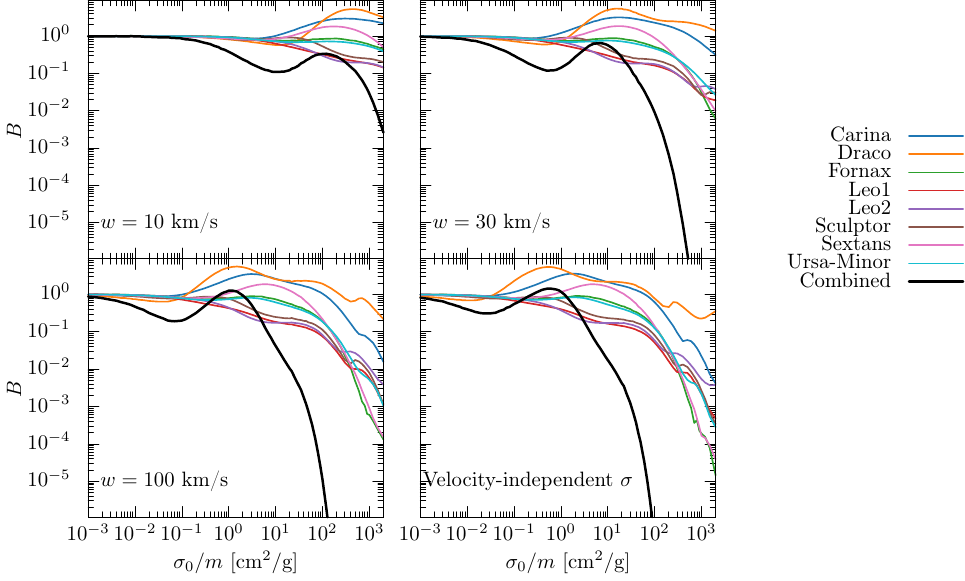}
    \caption{The same as Fig.\,\ref{fig:UFD} but for the classical dSphs.}
    \label{fig:classical}
\end{figure}

In Fig.\,\ref{fig:classical}, 
we show the Bayes factors for the classical dSphs.
They exhibit a gradual decline in the Bayes factor with increasing $\sigma_0/m$.
This suggests that gravothermal core collapse plays a less significant role in shaping the likelihoods for these galaxies. 


Figure\,\ref{fig:2D} presents the results combining the ultrafaint and classical dSphs.
For $w \gtrsim 30\,\mathrm{km/s}$, the cross-section region of $\sigma_0/m \gtrsim 0.2\,\csunit$ is decisively disfavored compared to CDM.
For $w<\order{10}\,\mathrm{km/s}$, on the other hand, this limit weakens to $\sigma_0/m \lesssim 10\,\csunit$ and a region around very high cross-section values, $\sigma_0/m \sim 10^3\,\csunit$, shows increased evidence.

The model evidence can be interpreted as a conventional Bayesian parameter constraint. Assuming a log-uniform prior on $\sigma_0/m$ over [$10^{-3},10^3$]\,$\csunit$, we obtain 95\%-percentile upper limits of $0.8$\,$\csunit$ for $w=10$\,km/s and $0.04$\,$\csunit$ for the velocity-independent case.

\begin{figure}[t]
    \centering    
    \includegraphics[width=\linewidth]{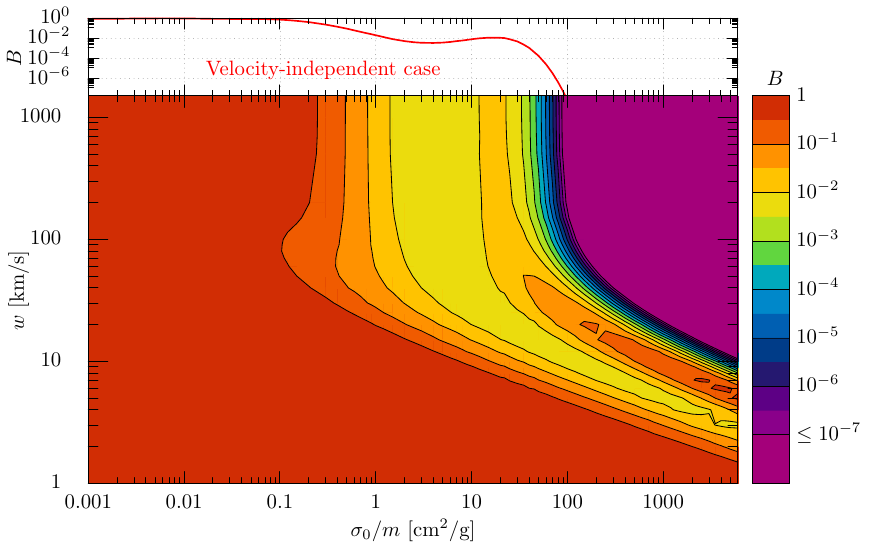}
    \caption{The same as Fig~\ref{fig:2D}, but the case of $V_{50}=18.0$~km/s for UFDs.
    }
    \label{fig:2D_v18}
\end{figure}

Motivated by galaxy formation theory at the reionization epoch~\cite{Fitts:2018eww}, $V_{50}=18$\,km/s for UFDs are another choice for this parameter. 
In this case, the Bayes factor result improves for the SIDM models (Fig.\,\ref{fig:2D_v18}).
However, recent studies~\cite{2019MNRAS.488.4585G,DES:2019ltu} found that $V_{50}=18$\,km/s is disfavored by the abundance of Milky Way satellites. Therefore, we adopt $V_{50}=10.5$\,km/s as the fiducial value.

\textbf{\textit{Discussion}}---Our dynamical analyses do not account for baryonic feedback effects. 
However, SIDM + hydrodynamical simulations~\citep{2017MNRAS.472.2945R,Straight:2025udg} 
show feedback effects can induce core formation even for $\sigma_0/m = 1\,\csunit$ 
in the classical dSph mass regime ($M_\ast \sim 10^6\text{--}10^7 M_\odot$). 
In contrast, UFDs are largely unaffected, so Bayes factors while the Bayes factor for the classical dSphs 
might be altered, especially at lower $\sigma_0/m$.

The current SASHIMI framework also lacks spatial information, using an average 
subhalo distribution over the host halo's virial volume. Adding Gaia proper motions information
enables estimates of infall times and pericenters~\citep{2019ApJ...875...77P, 
Li:2021nuw, 2022A&A...657A..54B}, offering insight into each dwarf's mass loss. 
This is a potential improvement for future work. However, as we combine multiple 
satellites, individual variations are mostly averaged out.

The present model also assumes spherical symmetry and equilibrium, yet real dSphs may be 
non-spherical and tidally disturbed. Still, SIDM studies typically rely on these 
assumptions. Extending SIDM to non-spherical, non-equilibrium cases is important 
future work.

The satellite galaxies Antlia~2, Crater~2, and Tucana~3 are excluded in our analysis. 
Antlia~2 and Crater~2 are 
extremely diffuse~\citep{2016MNRAS.459.2370T,2019MNRAS.488.2743T} with very low 
velocity dispersions~\citep{2017ApJ...839...20C,2021ApJ...921...32J}, implying 
underdense halos due to strong tides~\citep{2021ApJ...921...32J,2021MNRAS.504.2868P,
2020arXiv200606681S}. Tucana~3 has likely undergone strong disruption due to its 
close pericenter radius~\citep{2018ApJ...866...22L,2022A&A...657A..54B}. SASHIMI 
targets typical subhalo evolution, making it difficult to capture such outliers.

\textbf{\textit{Conclusions}}---In this work, we extended the SIDM analysis of UFDs presented in Ref.\,\cite{Hayashi:2020syu} by incorporating the effects of gravothermal core collapse using the SASHIMI-SIDM framework. Furthermore, we expanded our analysis to include updated kinematics data for both classical and ultrafaint dSphs.

Our findings highlight the interplay between core formation and gravothermal collapse, which manifests in distinct trends in the Bayes factors across different cross-section parameters. In particular, UFDs exhibit a bimodal preference, favoring both small and large self-interaction cross sections, whereas classical dSphs show a more gradual decline in statistical preference as the cross section increases. 
The combined constraints from these systems provide no substantial evidence favoring SIDM over the CDM paradigm.
Combining with the constraints from galaxy clusters~\cite{Andrade:2020lqq}, our result plays a complementary and interconnected role in advancing our understanding of dark matter.

\begin{acknowledgments}
\textit{Acknowledgments}--- This work is supported by Grant-in-Aid for Scientific Research from the Ministry of Education, Culture, Sports, Science, and Technology (MEXT), Japan, grant numbers 20H05850, 24K07039 (S.A.), JP20H05861 (S.A. and S.H.), 23K13098 (S.H.),
 20H01895, 21K13909, 23H04009, JP24K00669 (K.H.),
 21H04471, 22K03615, 24K23938 (M.I.), 
 20H01895 and 20H05860  (S.S.) and by World Premier International Research Center Initiative (WPI), MEXT, Japan. 
 The work of S.S. is supported by DAIKO FOUNDATION.
\end{acknowledgments}

\bibliography{main}

\clearpage

\textbf{
\textit{Appendix A: Effective constant cross section}}---Regarding the cross section, the analysis in Refs.\,\cite{Yang:2022hkm,Yang:2022zkd,Yang:2022hkm} demonstrated that the halo evolution in SIDM models with velocity- and angular-dependent cross sections can be effectively captured by an effective constant cross section, 
\begin{align}
\label{eq:sigmaeff}
    \sigma_{\mathrm{eff}} &= 
\frac{
    2 \int dv\, d\cos\theta \frac{d\sigma}{d\cos\theta} \sin^2\theta\, v^5 f_{\mathrm{MB}}(v, v_{\mathrm{eff}})
}{
    \int dv\, d\cos\theta \sin^2\theta\, v^5 f_{\mathrm{MB}}(v, v_{\mathrm{eff}})
}\ , \\
&= \sigma_0 \mathcal{F}(w^2/v_\mathrm{eff}^2)\ , \\
\mathcal{F}(x)&=-\frac{x^2}{64} \qty(4 + e^{x/4}(4+x)\mathrm{Ei}(-x/4)
) \ .
\end{align}
Here, $\mathrm{Ei}(x)$ is the exponential integral function, $\mathrm{Ei}(x) =
-\int_{-x}^\infty dt e^{-t}/t$.
In Eq.\,\eqref{eq:sigmaeff},
$f_\mathrm{MB}(v,v_\mathrm{eff})$ denotes
the one-dimensional Maxwell-Boltzmann distribution,
\begin{align}
    f_\mathrm{MB}\propto v^2 e^{-v^2/(4v_\mathrm{eff}^2)} \ ,
\end{align}
with $v_\mathrm{eff} = 0.64 \hat{V}_\mathrm{max}$.
We have neglected the dump of the velocity distribution due to the escaping.
In the limit of $w\to \infty$,
the cross section becomes velocity-independent and 
$\sigma_\mathrm{eff}\to \sigma_0$.

\vspace{5pt}
\textbf{
\textit{Appendix B: Stellar density and velocity anisotropy profiles}}---For the stellar density profile, we adopt the Plummer profile,
$\nu(r)\propto (1+r^2/r_{1/2}^2)^{-5/2}$,
implying surface density profile of the form
$\Sigma(R)\propto (1+R^2/r_{1/2}^2)^{-2}$,
where $r_{1/2}$ is the half-light radius which is one of the free parameters in this work.

In our analysis, we assume the following anisotropy profile, which holds a high degree of generality,
\begin{align}
\label{eq:anisotropy}
\beta_{\mathrm{ani}}(r) = \frac{\beta_0+\beta_{\infty}(r/r_{\beta})^{\eta}}{1+(r/r_{\beta})^{\eta}}\ .
\end{align}
The four parameters are the inner anisotropy $\beta_0$, outer anisotropy $\beta_{\infty}$, sharpness of the transition $\eta$, and transition radius $r_{\beta}$.

The prior distributions of the astrophysical parameters are assumed to follow uniform distributions within the following ranges:
$0 \le \log_{10}(r_{\beta}/{\rm pc}) \le 4$, $1 \le \eta \le 10$, $0 \le 2^{\beta_{0}} \le 1$ and $0 \le 2^{\beta_{\infty}} \le 2$.

\vspace{5pt}
\textbf{
\textit{Appendix C: The line-of-sight velocity dispersion profiles}}---We compare the line-of-sight velocity dispersion profiles obtained from the SASHIMI-SIDM priors~(for $\sigma_0/m=\{0,10,100\}$\,cm$^{2}$/g) with the binned observational data for Segue~1 in Fig.\,\ref{fig:sigmalos_seg1}, as well as for Draco and Fornax in Fig.\,\ref{fig:sigmalos_classical}.
For Segue~1, we compute the dispersion profiles for the $V_{50}=10.5$\,km/s and $18.0$\,km/s cases motivated by the formation criteria of UFDs.
We select these dSphs because the Bayes factor for Segue~1 exhibits a prominent bimodal feature in Fig.\,\ref{fig:UFD}, while Draco favors and Fornax disfavors SIDM models at larger $\sigma_0/m$ in Fig.\,\ref{fig:classical}.
In these figures, the prior with $\sigma_0/m = 100$\,cm$^{2}$/g yields a broader and higher $\sigma_\mathrm{los}$ profile than the others because the gravothermal collapse happens around this $\sigma_0/m$ regime.
While this prior can reproduce the observed $\sigma_\mathrm{los}$ for Segue~1 and Draco, it overestimates the observed $\sigma_\mathrm{los}$ in the center of Fornax.

\vspace{5pt}
\textbf{
\textit{Appendix D: The prior distributions from SASHIMI-SIDM}}---Figure\,\ref{fig:prior} displays an example of the corner plots of the predicted distributions due to satellite prior of the SIDM parameters at present, $P_{\rm sat}(\rho_s,r_s,r_c)$, for a given nominal cross section with $w=30$\,km/s.
It is clear from this figure that a bimodal distributions of each parameter at $\sigma_0=500$~cm$^2$/g, because of gravothermal collapse.

\vspace{5pt}
\textbf{\textit{Appendix E: The Bayes factor for the $\boldsymbol{V_{50}=18}$\,km/s case}}---Figure\,\ref{fig:2D_v18} displays the contour plot of the Bayes factor for the $V_{50}=18.0$~km/s case in UFDs.
Comparing with Fig.\,\ref{fig:2D}, the region with $\log_{10}B\leq-2$ is shifted toward larger $\sigma_0/m$. Assuming a velocity-independent cross section, this analysis favors cold dark matter (CDM) over self-interacting dark matter (SIDM) when $\sigma_0/m\gtrsim 1.0\,\csunit$.

\end{document}